\documentclass[twocolumn,showpacs,aps,prb,amsmath,amssymb,floatfix]{revtex4}
\usepackage{graphicx}
\usepackage{epstopdf}
\usepackage{dcolumn}
\usepackage{bm}
\usepackage{amsmath}
\begin{document}
\title{Topological superconductivity and fractional Josephson effect 
in quasi-one dimensional wires on a plane}
\author{E. Nakhmedov$^{1,2}$, S. Mammadova$^{1}$, and O. Alekperov$^{1}$}
\affiliation{$^1$Institute of Physics, Azerbaijan National Academy of
Sciences, H. Cavid ave. 33, AZ1143 Baku, Azerbaijan\\
 $^2$Institut f\"ur Theoretische Physik, Universit\"at W\"urzburg,
D-97074 W\"urzburg, Germany}
\date{\today}
\begin{abstract}
A time-reversal invariant topological superconductivity is suggested to be realized in a quasi-one
dimensional structure on a plane, which is fabricated by filling the superconducting 
materials into the periodic channel 
of dielectric matrices like zeolite and asbestos under high pressure. The topological superconducting phase
sets up in the presence of large spin-orbit interactions when intra-wire s-wave and inter-wire 
d-wave pairings take place. 
Kramers pairs of Majorana bound states emerge at the edges of each wire. We analyze effects of Zeeman magnetic
field on Majorana zero-energy states. In-plane magnetic field was shown to make asymmetric the energy dispersion,
nevertheless Majorana fermions survive due to protection of a particle-hole symmetry. 
Tunneling of Majorana quasi-particle from the end of one wire to the
nearest-neighboring one yields edge fractional Josephson current with $4\pi$-periodicity.  
\end{abstract}
\maketitle
\emph{Introduction}-The recent theoretical prediction and experimental observation of topological phase of matter
with time-reversal symmetry in a number of materials \cite{hk10} have attracted great interest
in this subject. The time-reversal invariant (TRI) topological superconductors (SCs) was predicted 
by theoretical classification of Bogolyubov-de Gennes (BdG) Hamiltonian \cite{srfl08, qhz10, tk10} which 
constitute a completely distinct symmetry class {\it DIII}. 
Due to the presence of intrinsic particle-hole symmetry (PHS) the gappless zero-modes
in the topological superconductors constitute Majorana fermions (MFs), obeying non-Abelian braiding
statistics which is useful in implementing fault-tolerant topological quantum computer \cite{nssf08}. 
A variety of condensed- matter systems hosting localized Majorana quasi-particles have been 
proposed, notably quantum Hall states \cite{mr91} and topological superconductors \cite{fk08, stf09, slts10,
lss10, oro10}, as well as systems with the charge-density wave instability \cite{na14}.

Recent works \cite{zkm13, zkmm13, wl12, dvo12, kfsb13, gpf14} have proposed TRI topological 
superconductivity (class $DIII$) with a $\mathbb{Z}_2$ invariant, which takes a value $\nu =1$,
in a number of systems with intrinsic or proximity induced superconductivity of {\it p}-wave, 
spin-triplet or $d$-wave and $s_{\pm}$-wave spin-singlet pairings. TRI topological SC is assumed 
to be realized in the variety class of natural quasi-one dimensional (quasi-1D) materials such  
as Lithium molybdenum purple bronze ($Li_{0.9}Mo_6O_{17}$) and some organic superconductors 
\cite{mbxw12, ls13, lbcs01}.

A possible application of Majorana fermions as topological qubits requires, on the one hand, finding
unique fingerprints that unambiguously confirm their existence and, on the other hand, developing
techniques that allow their detection. The majority of  previous proposals for their detection rely
on the zero-energy excitation features in the tunneling experiments. Alternative experimental
routes to observe Majorana bound states may be based on unconventional $4 \pi$-periodic
oscillation of the Josephson current. It has first been shown by Kitaev \cite{kitaev01} for two
single-channel topological superconducting wires, brought into contact to form SNS Josephson
junction, that the two Majorana edge states across the junction couple to each other and generate
a fractional Josephson effect: instead of the usual $2\pi$, the Josephson current exhibits
a periodicity of $4\pi$ with the phase difference between the superconductors. 
This doubling of the periodicity can be interpreted as the tunneling of ``half'' of a Cooper pair. 
This prediction has later been extended to many different systems \cite{fk09, jpar11}. 

Nowadays, the challenge is to find a real material which supports the topological SC properties. In this work 
we show that the time-reversal invariant topological superconducting phase can be realized in 
quasi-1D wires on a plane in the presence of s-wave intra-wire and  d-wave 
inter-wire pairings. Similar structures have been fabricated by Bogomolov's group \cite{bogom71, bk75} by filling a 
superconducting material into the cavities or channels of dielectric matrices like 
zeolite and asbestos crystals under high pressure up to $30~kbar$. The regular set of channels or cavities 
of $5-10 \AA$ diameter in zeolite and from $20-30 \AA$ up to $100-150 \AA$ diameter in asbestos \cite{bogom78}
form a periodic lattice of different geometry in one-, two- and three-dimension, e.g. like several equidistant 
filaments with $5 \div 20 \AA$ separation in zeolite and  with $150 \div 500 \AA$ separation in asbestos 
on a plane or quasi-1D space lattice.   
The critical temperatures $T_c$ of such structures become higher than the $T_c$ of the bulk 
superconductors by factors of $2-5$  \cite{bogom71}. High stress field around the filaments may
guarantee higher value of spin-orbit interactions in the structures. An increase in $T_c$ 
may be caused by an excitonic mechanism of inter-wire pairings due to polarization of the dielectric matrix between 
the wires. We investigate topological phases of such kind quasi-1D superconductor with combined 
s- and d-wave pairings in the presence and absence of the time-reversal invariance. 
In difference from chiral superconductors, belonging to $DIII$ symmetry-class too, the zero mode in 
the TRI topological superconductors come in pairs due to
Kramers's theorem. Multiple Majorana-Kramers pairs with strongly spatial overlapping wave functions 
are protected by time-reversal symmetry, and they persist at zero energy. 
At each end of a superconducting wire are localized two Majorana fermions that form
a Kramers doublet and are protected by time-reversal symmetry. 
An external Zeeman magnetic field breaks the Kramers degeneracy. Calculations of the energy dispersion of the
topological SC reveal two kind asymmetries: first, an interplay of Rashba and Dresselhaus SOIs makes the energy
dispersion strongly asymmetric even in the absence of the magnetic field, and second, in-plane Zeeman field 
introduces an additional anisotropy into the dispersion in the presence of Rashba or/and Dresselhaus SOIs even for a
single wire when $t_{\perp}=0$. In the non-trivial topological phase, Majorana
particles, resided at the ends of each wire, tunnel from one wire to the nearest-neighboring wire yielding 
fractional Josephson current over the wires' end with $4 \pi$ periodicity. 
  
\begin{figure}[th]
\[
\begin{tabular}{cc}
\includegraphics[scale=0.16]{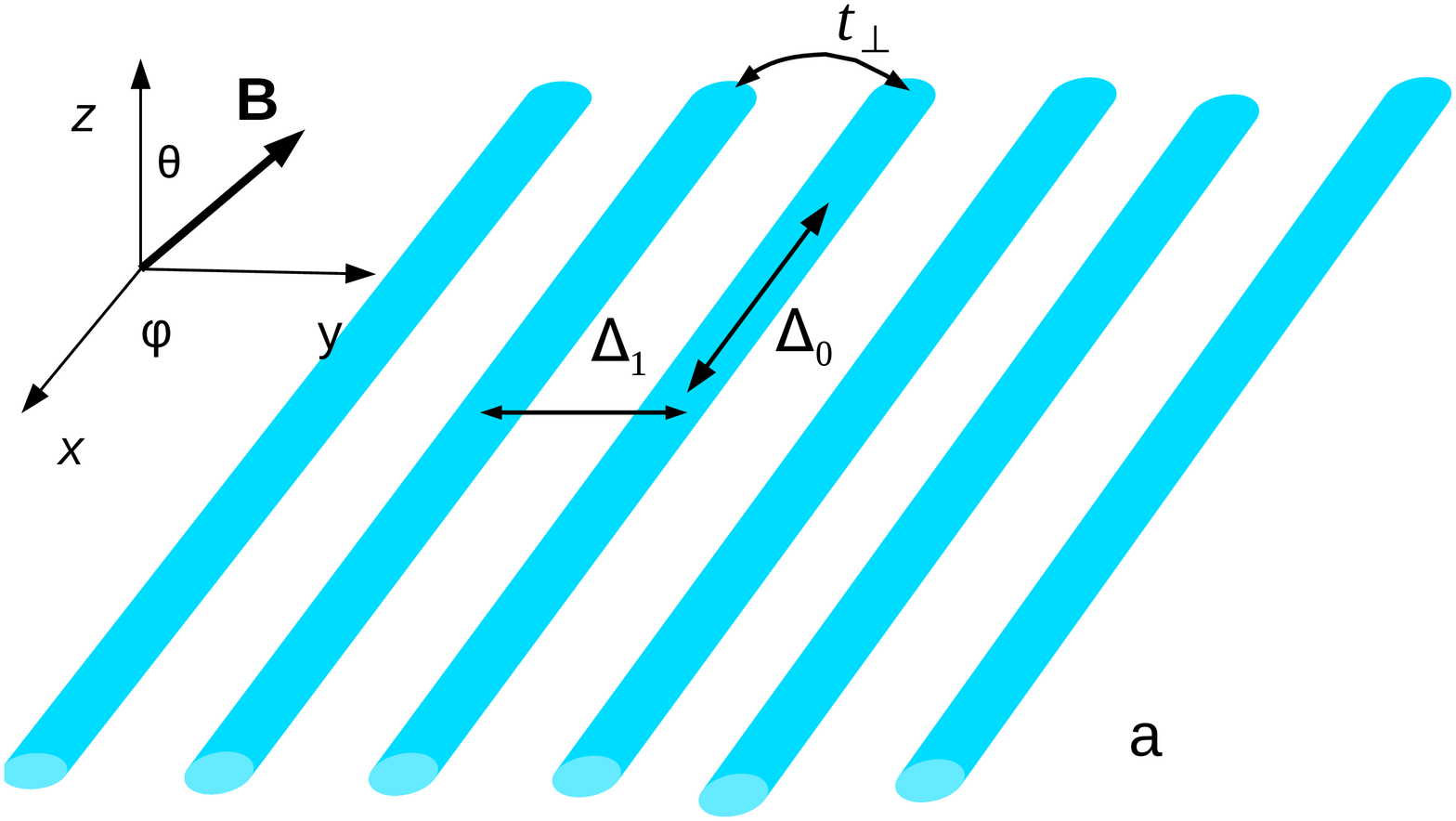} & \includegraphics[scale=0.45]{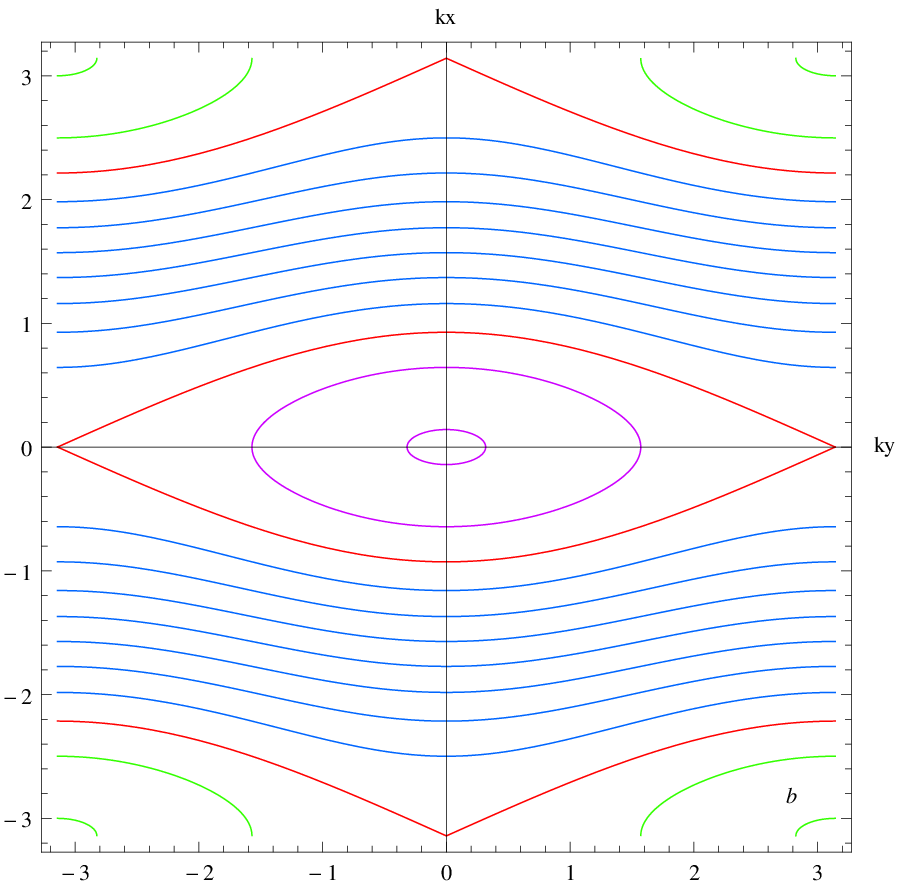} \\[10mm]
\includegraphics[scale=0.45]{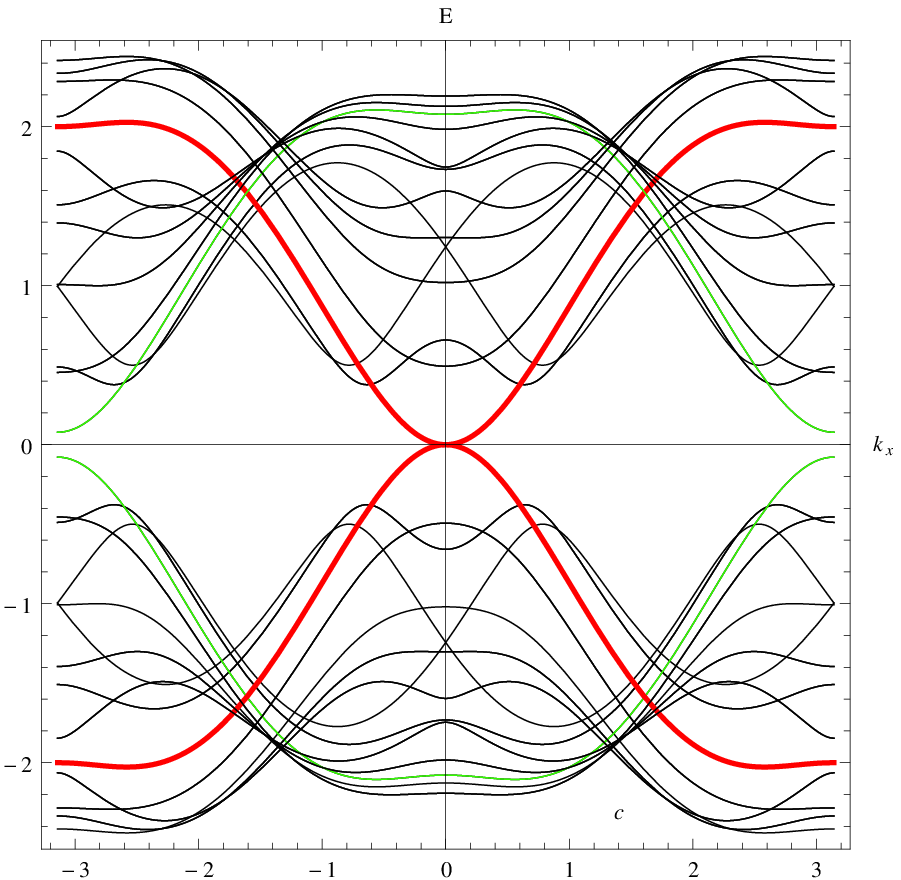} & \includegraphics[scale=0.45]{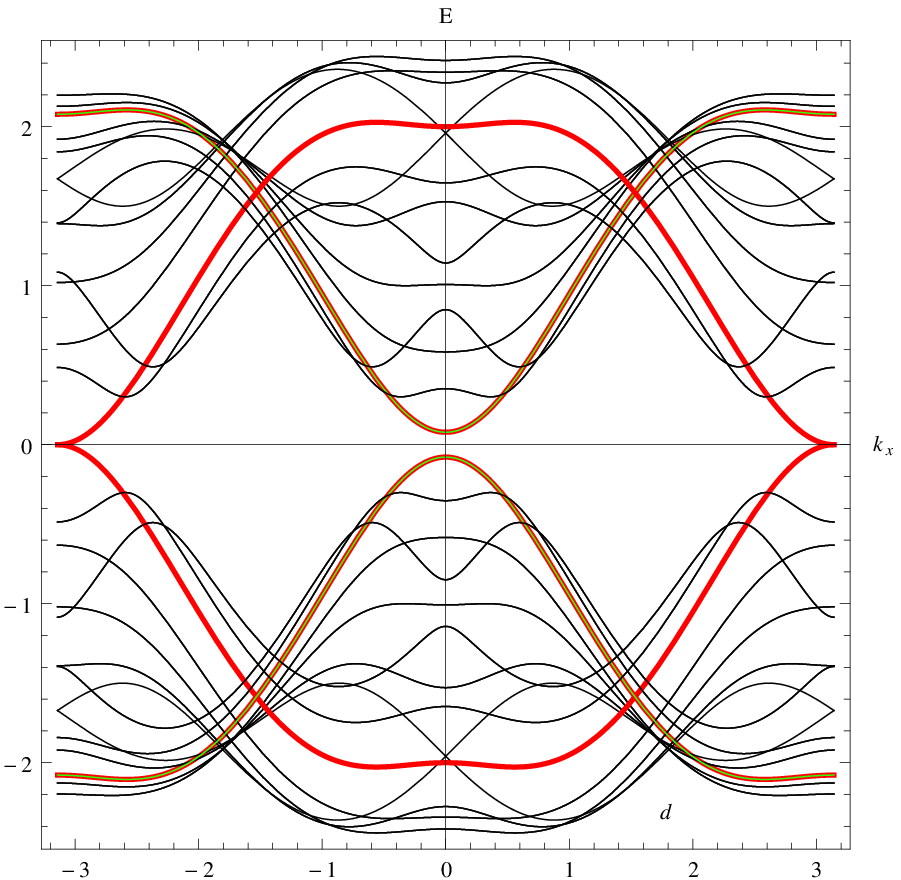}  \\
\end{tabular}
\]
\caption{ (a) Structure of quasi-1D superconducting wires on a plane with arbitrary 
directed magnetic field ${\bf B}$. 
(b) Fermi surface of the equidistant wires on plane, $-2 t_{\|} \cos k_x - 2 t_{\perp} \cos k_y = \mu$. 
Strong anisotropy is provided by the parameter $t \equiv t_{\perp}/t_{\|}=0.2$. The Fermi energy $\mu$
varies in the interval $|\mu| \le 2 t_{\|}+ 2 t_{\perp}$. The Fermi surfaces are opened for 
$|\mu| \le 2t_{\|}- 2 t_{\perp}$, which are shown by blue curves. For 
$2t_{\|}-2t_{\perp} < |\mu| \le 2t_{\|}+2t_{\perp}$ the Fermi surfaces are closed (green curves).
The red curves, corresponding to $|\mu|=2t_{\|}-2t_{\perp}$, separate the opened Fermi surfaces from closed ones.
The energy spectrum of strongly anisotropic 2D superconductor as a function of $k_x$ for different values of
$k_y= 0.0,~\pm 0.5,~ \pm \pi/3, ~\pm 1.5, ~\pm 2\pi/3,~ \pm 2.5$ and $ \pm 3.0$ 
for (c) $\Delta_0 >0$ with a zero-energy state (red curves) at the center of the Brillouin zone  and (d) $\Delta_0<0$ 
with a zero energy state (red curves) at the boundaries of the Brillouin zone. The parameters are chosen
in the unit of the band width $2t_{\|}=1$. In the both cases we choose
$|\Delta_0|=\Delta_1= 0.5 \times 2t_{\|}$, $t_{\perp}=0.2 \times 2t_{\|}$, $\alpha= 0.6 \times 2t_{\|}$, and $\beta=0$, also
$\mu= -0.061 \times 2t_{\|}$ and $\mu=0.061 \times 2t_{\|}$ for (c) and (d) correspondingly.}  
\label{fig1}
\end{figure}

\emph{Time reversal invariant topological superconductor}--~~
The equidistant superconducting wires, aligned along $x$-axes in $\{x, y\}$ plane, with $s$-wave intra- and 
$d$-wave inter-wire pairings (see, Fig. \ref{fig1}a),  in the presence of spin-orbit interactions 
(SOIs) and arbitrary directed homogeneous magnetic field ${\bf B}$ are described by Hamiltonian
\begin{equation}
\hat{H}=\sum_j\left\{\hat{H}_{j,j}+\hat{H}_{j,j+1}+\hat{H}_{j+1,j}\right\},
\label{H-main}
\end{equation}
where $\hat{H}_{j+1,j}=\hat{H}_{j,j+1}^{\dag}$, and
\begin{eqnarray}
&&\hspace{-5mm}\hat{H}_{j,j}=\sum_{\sigma, \sigma'}\int \frac{dk_x}{2\pi} \Big\{ \psi_{j, \sigma}^{\dag}(k_x) \xi_{k_x}
\psi_{j, \sigma}(k_x)+\nonumber\\
&&\hspace{-5mm} \psi_{j,\sigma}^{\dag}\Big(2\sin k_x \left[\alpha (\sigma_y)_{\sigma, \sigma'} +
\beta (\sigma_x)_{\sigma, \sigma'} \right]+\epsilon_Z \big[(\sigma_z)_{\sigma,\sigma'} \cos \theta+\nonumber\\ 
&&\hspace{-5mm}(\sigma_x)_{\sigma,\sigma'} \sin \theta~\cos \varphi +
(\sigma_y)_{\sigma,\sigma'} \sin \theta \sin \varphi \big]\Big)\psi_{j,\sigma'}+\nonumber\\
&&\hspace{-5mm}\Delta_0 \psi_{j,\uparrow}^{\dag}(k_x)\psi_{j,\downarrow}^{\dag}(-k_x)+ 
\Delta_0^{\ast}\psi_{j, \downarrow}(-k_x) \psi_{j, \uparrow}(k_x)\Big\},
\label{H-dia}
\end{eqnarray}
\begin{eqnarray}
&&\hspace{-5mm}\hat{H}_{j,j+1}=\sum_{\sigma, \sigma'}\int \frac{dk_x}{2\pi}\Big\{ t_{\perp}
\psi_{j,\sigma}^{\dag}(k_x)\psi_{j+1,\sigma}(k_x) +\nonumber\\
&&\hspace{-5mm} i \psi_{j,\sigma}^{\dag} \big[\bar{\alpha}
(\sigma_x)_{\sigma,\sigma'} + \bar{\beta}(\sigma_y)_{\sigma, \sigma'}\big]
\psi_{j+1,\sigma'} +\nonumber\\
&&\hspace{-5mm}\Delta_1 \psi_{j,\uparrow}^{\dag}(k_x)\psi_{j+1,\downarrow}^{\dag}(-k_x)+\Delta_1^{\ast} \psi_{j,\downarrow}(-k_x)
\psi_{j+1,\uparrow}(k_x)  \Big\}.
\label{H-off}
\end{eqnarray}
Here, $\psi_{j,\sigma}^{\dag}(k_x)$($\psi_{j,\sigma}(k_x)$) is a creation (annihilation) operator of an electron
with a longitudinal momentum $k_x$ and spin $\sigma$ in a $j$th wire, $t_{\|}$ ($t_{\perp}$) and $\mu$ are the
longitudinal (transverse) overlap integral and the Fermi energy, $\xi_{k_x}=(-2 t_{\|}\cos k_x -\mu)$ 
is the energy dispersion in a single wire, $\alpha (\beta)$ and $\bar{\alpha} 
(\bar{\beta})$ are the longitudinal and transverse components of Rashba (Dresselhaus) spin-orbit constant; 
$\epsilon_Z=g \hbar \mu_B B/2$ is the Zeeman energy,
$\theta$ and $\varphi$ are correspondingly the polar and the azimuthal angles of the 
magnetic field ${\bf B}=B\{\sin \varphi \cos \theta,~ \sin \varphi \sin\theta,~\cos \theta\}$. 
Note that the orbital effects of the magnetic field, which is neglected here, will be considered elsewhere.
The existence of intra- and inter-wire order parameters $\Delta_0$ and $\Delta_1$ allows one
to introduce the effective order parameter $\Delta(k_y)$ in momentum space, 
$\Delta (k_y)=\Delta_0 +2 \Delta_1 \cos k_y=|\Delta (k_y)|e^{i\phi}$, \cite{nt94}.
 Further, we will take $\alpha = \bar{\alpha}$ and $\beta = \bar{\beta}$. The structure is
strongly anisotropic under $t_{\perp} \ll t_{\|}$, and the Fermi surface is opened (see, Fig. \ref{fig1}b), 
consisting of two goffered lines. Note that a conventional 
Josephson coupling between the wires is realized  under $t_{\perp} < k_BT_{c0} \ll 2t_{\|}$,
\cite{el74}, where $T_{c0}$ is the critical temperature evaluated by means of the mean field theory.

We write Hamiltonian, given by Eqs. (\ref{H-main})-(\ref{H-off}), as $N \times N$ tridiagonal matrix 
$\hat{H}=\int \frac{dk_x}{2\pi}\Psi_N^{\dag}\mathcal{H}_N \Psi_N$ in the 
basis of the generalized Nambu wave function of $N$ superconducting wires,
$\Psi_N^{\dag}=(\Psi_{1,k_x}^{\dag}, \Psi_{2,k_x}^{\dag} \cdots \Psi_{j,k_x}^{\dag} \cdots \Psi_{N,k_x}^{\dag})$, 
which is $4N$ dimensional vector with $\Psi_{j,k_x}^{\dag}=\left(\psi_{j,\uparrow}^\dag(k_x), \psi_{j,\downarrow}^{\dag}(k_x),
\psi_{j,\downarrow}(-k_x),-\psi_{j, \uparrow}(-k_x)\right)$ for a single wire.
This expression can be exactly written in the two-wire Nambu basis, 
$\Psi_j^{(2)~\dag}=(\Psi_{j,k_x}^{\dag}, \Psi_{j+1,k_x}^{\dag})$,
\begin{equation}
\hat{H}=\int \frac{dk_x}{2\pi} \sum_{j=1}^{N}\Psi_j^{(2)~\dag}\mathcal{H}_j^{(2)}\Psi_j^{(2)},
\label{H-2}
\end{equation}   
where 
\begin{eqnarray}
\mathcal{H}_j^{(2)}=
\left( \begin{array}{cc}
\mathcal{H}_{j,j} & \mathcal{H}_{j,j+1}\\
\mathcal{H}_{j+1,j} & \mathcal{H}_{j+1,j+1}
\end{array} \right),
\label{m-M}
\end{eqnarray}
and each $\mathcal{H}_{j,j'}$ entry is $4\times 4$ matrix,
\begin{eqnarray}
&&\mathcal{H}_{j,j}= \frac{1}{2}\Big\{\xi_{k_x} \tau_z +2 \alpha \sin k_x  \tau_z \sigma_y + 
2 \beta \sin k_x \tau_z \sigma_x +\nonumber\\ 
&&\epsilon_Z(\cos \theta  \sigma_z +
\sin \theta \cos \varphi \sigma_x +\nonumber\\
&& \sin \theta \sin \varphi \sigma_y)+
|\Delta_0| \cos \phi_{0j}  \tau_x - |\Delta_0| \sin \phi_{0j} \tau_y\Big\},
\label{H-jj}
\end{eqnarray}
and $\mathcal{H}_{j+1,j}=\mathcal{H}_{j,j+1}^{\dag}$,
\begin{eqnarray}
&&\mathcal{H}_{j,j+1}= \frac{1}{2}\Big\{t_{\perp} \tau_z +i\bar{\alpha} \tau_z \sigma_x+
i \bar{\beta} \tau_z \sigma_y+\nonumber\\
&&|\Delta_1| \cos \phi_{1j} \tau_x -|\Delta_1| \sin \phi_{1j} \tau_y \Big\}.
\label{H-jj+1}
\end{eqnarray}
Note that $\mathcal{H}_{N, N+1}=\mathcal{H}_{N+1,N}=0$ and  $\mathcal{H}_{N+1, N+1}=\mathcal{H}_{1,1}$ for 
$\mathcal{H}_j^{(2)}$ and $\Psi_N^{(2)~\dag}=(\Psi_{N,k_x}^{\dag}, \Psi_{1,k_x}^{\dag})$ in Eq. (\ref{m-M}). 
Hamiltonian, given by Eqs. (\ref{H-main})-(\ref{H-off}), is transformed from 
discrete wire-number representation to the momentum space, and is written in the extended Nambu spinor 
basis $\Psi_{\bf k}^{\dag}=(\psi_{{\bf k}, \uparrow}^{\dag},\psi_{{\bf k}, \downarrow}^{\dag}, \psi_{{-\bf k}, \downarrow}, 
-\psi_{{-\bf k}, \uparrow})$, as
\begin{equation}
\hat{H}=\int_{-\pi}^{\pi}\frac{dk_x}{2\pi}\int_{-\pi}^{\pi}\frac{dk_y}{2\pi}\Psi_{\bf k}^{\dag} \mathcal{H}\Psi_{\bf k},
\label{H-int}
\end{equation}
where
\begin{eqnarray}
&&\mathcal{H}= \xi_{\bf k} \tau_z +2 (\alpha \sin k_x -\bar{\beta} \sin k_y) \tau_z \sigma_y + 2 (\beta \sin k_x -\nonumber\\ 
&&\bar{\alpha} \sin k_y) \tau_z \sigma_x + \epsilon_Z(\cos \theta \sigma_z +
\sin \theta \cos \varphi \sigma_x +\nonumber\\
&&\sin \theta \sin \varphi \sigma_y)+
|\Delta| \cos \phi ~ \tau_x - |\Delta| \sin \phi~ \tau_y,
\label{H-nambu}
\end{eqnarray}
where $\xi_{\bf k}=-2t_{\|} \cos k_x-2t_{\perp} \cos k_y -\mu$, and $\phi$ is the total phase of the effective order
parameter $\Delta (k_y)$. 

In the absence of the magnetic field and for real order parameter ($\phi=0$), the time-reversal symmetry 
and the particle-hole symmetry are protected.  Hamiltonian $\mathcal{H} (B=0) \equiv \mathcal{H}_0$ 
satisfies the relations  $\Theta \mathcal{H}_0({\bf k})= \mathcal{H}_0(-{\bf k})\Theta$ with
TRI operator $\Theta = i \sigma_y \mathcal{K}$, where $\mathcal{K}$ is the anti-unitary complex conjugation
operator.  The PHS emerges from the intrinsic structure of the BdG Hamiltonian, which 
relates quasi-particle excitations at $\pm E$ through the second quantized operator relation 
$\gamma_E=\gamma_{-E}^{\dag}$, providing a ground for formation of zero-mode Majorana state. This symmetry satisfies
the anti-commutation relation
$\Xi \mathcal{H}_0({\bf k})= - \mathcal{H}_0(-{\bf k}) \Xi$ with the particle-hole operator 
$\Xi= \tau_y \sigma_y \mathcal{K}$, obeying $\Xi^2=-1$. The presence of TRS and PHS leads to a chiral 
symmetry $\Pi \mathcal{H}_0({\bf k})=-\mathcal{H}_0({\bf k}) \Pi$, 
where the unitary chiral operator $\Pi$ is the product of $\Theta$ and $\Xi$,
$\Pi=-i\Theta \cdot \Xi= \tau_y \sigma_0$. 

The energy spectrum, obtained from $det|E_0-\mathcal{H}_0|=0$, reads
\begin{equation}
E_0= s \sqrt{(\xi_{\bf k}  \pm  \epsilon_S)^2 +\Delta^2(k_y)},
\end{equation}
where  $s=\pm$ and
$\epsilon_S= 2 \Big[\left(\sin^2k_x+ \sin^2k_y \right)(\alpha^2+\beta^2)- 4\alpha \beta 
\sin k_x \sin k_y \Big]^{1/2}$,
is the spin-orbit interactions energy. 
The condition  $\Delta(k_y)=0$ determines the nodal points of the order parameter. 
$\Delta(k_y)$ changes sign at 
$\cos k_y = -  \frac{\Delta_0}{2\Delta_1}$ if $|\Delta_0|<2 \Delta_1$ as moving along $k_y$
from $k_y=0$ to $k_y=\pi$. A nontrivial TRI superconductor with $\nu=1$ is realized if there are an odd number of Fermi
surfaces with a negative pairing order parameter \cite{qhz10}.  
At the nodal points 
\begin{equation}
E_{0N}=s \Big[2 t_{\|} \cos k_x + \mu-  t_{\perp}\frac{\Delta_0}{2 \Delta_1} \pm \epsilon_{SN}(k_x,k_{yN})\Big],
\label{Enodal}
\end{equation}
where $\delta =\sqrt{1- \frac{\Delta_0^2}{4 \Delta_1^2}}$; and $\epsilon_{SN}(k_x,k_{yN})= 2 \sqrt{\left(\sin^2 k_x + 
\delta^2\right)(\alpha^2+\beta^2)- 
4\alpha \beta \delta \sin k_x}$ is the value of $\epsilon_S$ at the nodal point 
$k_{yN}=\arccos \left(-\frac{\Delta_0}{2 \Delta_1}\right)$. 
$\epsilon_{SN}$ varies between the maximal $\epsilon_{SN}^{max}(\pm \pi/2,k_{yN}, \alpha, \beta)=
2 \sqrt{\alpha^2+\beta^2} \sqrt{1 + \delta^2 \mp \frac{4\alpha \beta}{\alpha^2+\beta^2}\delta}$ and minimal 
$\epsilon_{SN}^{min}(k_{x0}, k_{yN}, \alpha,\beta)=2 \delta\frac{(\alpha +\beta)|\alpha - \beta|}
{\sqrt{\alpha^2+\beta^2}}$ values.

The order parameter switches sign as the
nodal point is crossed. On the other hand, the SOIs split the Fermi surfaces. The splitted Fermi 
surfaces around the nodal points lie in the energetic interval of $\epsilon_{SN}(k_x, k_{yN})$
from each other. Non-trivial TRI topological phase with $\nu=1$ is realized when the maximal value of the 
kinetic energy term (the first three terms) in Eq. (\ref{Enodal}) is smaller than the minimal value of the 
SOIs mediated splitting energy $\epsilon_{SN}^{min}(k_x, k_{yN}, \alpha, \beta)$, 
$\Big|2t_{\|} + \mu- t_{\perp}\frac{\Delta_0}{\Delta_1}\Big|< \epsilon_{SN}^{min}(k_{x0}, \alpha, \beta)$.
The SC is fully gapped when
$\Big|2t_{\|} + \mu- t_{\perp}\frac{\Delta_0}{\Delta_1}\Big|> \epsilon_{SN}^{max}(k_{x0}, \alpha, \beta)$.
The calculation of the BdG quasi-particles' energy spectrum is simplified for $\beta =0$, for which 
$\epsilon_{SN}^{min}(k_{x0},\alpha, \beta)=2\alpha \delta$ and 
$\epsilon_{SN}^{max}(k_{x0},\alpha, \beta)=2\alpha \sqrt{1+ \delta^2}$. 
The band structure
of the topological SC with zero energy surface states for this case is drawn in Figs. \ref{fig1}c, d. For 
$\Delta_0 <0$, Majorana edge states appear  at $k_x=0$, which are shown in
Fig. \ref{fig1}c by red curves. The zero energy states move to the Brillouin zone boundaries, $k_x= \pm \pi $,
for $\Delta_0>0$ (see, Fig. \ref{fig1}d). The topologically non-trivial 
superconductor belongs to $DIII$ symmetry class with $\mathbb{Z}_2$ invariant, which takes a value $\nu=1$.
In this case unpaired MFs at each end of a single wire form topologically protected
Majorana Kramers pairs, yielding four zero-energy modes. 

Even though a pair of zero energy states
are localized at the same end, they are protected by time-reversal symmetry against hybridization preventing
to split them to finite energies.
 Braiding the end-states in these TS results
in an exchange of the Kramers pairs rather than isolated Majorana mode, which complicates an application
of the TRI topological superconductors to quantum computation. Although braiding of two Majoranas
in chiral topological superconductors yields Abelian operators, braiding of Majorana end states in
$DIII$-class topological superconductors was shown \cite{lwl14} to represent by non-Abelian operators due to
protection of the TRS.    
 
\emph{Effects of in-plane magnetic field}--~~
An external magnetic field destroys the time-reversal 
symmetry in the system.  The superconductor turns into topological state for an appropriate choice 
of the magnetic field and SOI strengths as well as the value of the chemical potential. The energy 
dispersion for the BdG quasi-particles in the presence of an external magnetic field is expressed according to 
Eq. (\ref{H-nambu}) as,  
\begin{eqnarray}
&&(E^2-\xi_{\bf k}^2-\epsilon_S^2-\epsilon_Z^2-|\Delta|^2)^2- 8 \xi_{\bf k}\epsilon_Z \sin \theta~ 
\Phi_{\bf k}(\varphi) E -\nonumber\\  
&&4 \xi_{\bf k}^2(\epsilon_S^2+\epsilon_Z^2) -
4(\epsilon_z \sin \theta \Phi_{\bf k})^2- 4|\Delta|^2 \epsilon_Z^2=0,
\label{E-total}
\end{eqnarray}
where $\Phi_{\bf k}(\alpha, \beta, \varphi)=\sin k_x (\alpha \sin \varphi + \beta \cos \varphi) - 
\sin k_y (\alpha \cos \varphi + \beta \sin \varphi)$ determines the azimuthal angle $\varphi$ dependence  
of the energy. $\Phi_{\bf k}(\alpha, \beta, \varphi)$ emerges due to interference between the co-planar vector fields,
such as the magnetic field and the spin-orbit interactions, and introduces a linear in $E$ term 
in Eq. (\ref{E-total}). This linear term destroys a symmetry
of the energy dispersion, and it vanishes for a magnetic field, 
perpendicular ($\theta=0$) to the superconducting plane, yielding from Eq. (\ref{E-total}) for the energy spectrum,
\begin{equation}
E^2= \xi_{\bf k}^2+ \epsilon_S^2+\epsilon_Z^2+|\Delta(k_y)|^2 \pm 2\sqrt{\xi_{\bf k}^2(\epsilon_S^2+\epsilon_Z^2)+
|\Delta|^2 \epsilon_Z^2},
\end{equation} 
which hosts zero energy state $E(0)=s \big|\epsilon_Z \pm \sqrt{\tilde{\mu}^2 +|\Delta(0)|^2}\big|$ 
at the center of the Brillouin zone. Here, $\tilde{\mu}=\mu+2t_{\|}+2t_{\perp}$ and $s=\pm$. This expression
shows that for $\epsilon_Z >\sqrt{\tilde{\mu}^2+ |\Delta|^2}$ the topological non-trivial phase is realized, 
where one Majorana bound-state resides at $k=0$;
while for $\epsilon_Z < \sqrt{\tilde{\mu}^2+|\Delta|^2}$ a topologically trivial gapped state takes place with 
one Majorana bound-state at the edges.
\begin{figure}
\resizebox{.48\textwidth}{!}{%
\includegraphics[width=1cm]{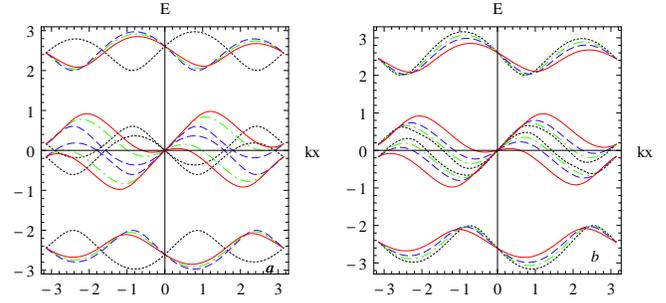}}
\caption {Energy dispersion vs $k_x$ ($k_y=0$) for (a) $\beta=0$ and different values of the 
in-plane magnetic 
orientation: $\varphi=3\pi/2,~2\pi/3,~\pi/2$ and $\pi/4$ for dotted (black), dot-dashed (green), 
dashed (blue) and  solid (red) curves, respectively; (b) $\varphi =\pi/4$ and different
values of Dresselhaus SOI constant: $\beta=0.4,~0.3, ~0.2$ and $0.0$  for dotted (black), dot-dashed (green), 
dashed (blue) and  solid (red) curves, respectively. The dimensionless parameters for both figures are
chosen (in the unit of $2t_{\|}$) to be $2t_{\perp}=0.1$, $\mu=0$, $\alpha=0.5$, $\epsilon_Z=\sqrt{1.7}$, and 
$|\Delta|=0.7$.}  
\label{fig2}
\end{figure}
The energy dispersion depends on orientation of the in-plane magnetic field ($\theta=\pi/2$), and the dependence 
of $E$ on $k_x$ and $k_y$ becomes asymmetric. It is well known \cite{sl03, nao11}, 
that interplay of Rashba- and Dresselhaus- SOIs in the absence of Zeeman magnetic field makes anisotropic the 
Fermi surface and the kinetic properties of $2D$ electron gas. In-plane magnetic field introduces an 
additional anisotropy in the energy dispersion. Note that the asymmetries introduced by these two factors
differ each other. Indeed, the anisotropy e.g. in the conductivity  was shown \cite{sl03} to increase with the strength
of SOIs (the case $\alpha = \pm \beta$ is particular \cite{sel03}). Nevertheless, the dispersion asymmetry 
oscillates with the in-plane magnetic field orientation. $E$ vs. $k_x$ dependence is 
drawn in Fig. \ref{fig2} for two different cases: the energy dispersion in Fig. \ref{fig2}a is calculated in the 
absence of the Dresselhaus SOI ($\beta=0$) and  for different orientations of the in-plane magnetic field; 
whereas Fig. \ref{fig2}b shows the dispersion for fixed value of the SOI constants ($\alpha=0.5\times 2t_{\|}$, 
$\beta=0.4\times 2t_{\|}$) but different values of the azimuthal angle $\varphi$. Symmetry of $E$ vs. $\{k_x, k_y\}$
in Eq. (\ref{E-total}) strongly differs for $B=0$ but $\beta \neq 0$ from $B \neq 0$ but $\beta =0$. 
Indeed, $E(-k_x,-k_y)= \pm E(k_x, k_y)$ in the former case declares the existence of both time-reversal and
particle-hole symmetries. Nevertheless, $E(-k_x,-k_y)= - E(k_x, k_y)$ in the latter case demonstrates that 
the particle-hole symmetry only is preserved in the system. Note that the energy dispersion becomes asymmetric
even for a single wire in the absence of the transverse tunneling $t_{\perp}=0$ (also $\bar \alpha =\bar \beta =0$),
when a linear in $E$ term in Eq. (\ref{E-total}) survives and becomes proportional to 
$\Phi_{\bf k}(\alpha, \beta, \varphi)=\sin k_x (\alpha \sin \varphi + \beta \cos \varphi)$. The chosen parameters allow a 
realization of the zero-energy state at the origin with asymmetric dispersion. While the Hamiltonian (\ref{H-nambu}) 
respects the particle-hole symmetry even in the presence of the magnetic field, a zero energy states obtained 
constitute Majorana fermions.   

The presence of the midgap mode alters the transport properties of the Josephson junction as the Majorana
edge states mediate the transfer of single electrons, as opposed to Cooper pairs, across the junction. Kitaev
first predicted \cite{kitaev01} that a pair of Majorana fermions fused across a junction of two 
topological superconducting
wires generates a Josephson current $I \propto \sin (\phi_l-\phi_r)/2$, exhibiting thus a remarkable $4\pi$ 
periodicity in the phase difference $\phi_l-\phi_r$ between the left and right wires, in contrast to the $2\pi$ 
periodic current in conventional Josephson contacts. Such a ``fractional'' Josephson effect was later established 
in other structures supporting Majorana edge states. 

In order to calculate the Majorana coupling between two nearest-neighboring wires, one needs to find the wave 
function of individual Majorana at the end of $j$th wire. A unitary transformation 
$\mathbb{H}_j= \mathbb{U}^{\dag} \mathcal{H}_j^{(2)} \mathbb{U}_j$ with the operator 
\begin{eqnarray}
\mathbb{U}_j=
\left( \begin{array}{cc}
 e^{-i \frac{\phi_{0j}}{2}\tau_z}& 0\\
0 & e^{-i \frac{\phi_{0j+1}}{2}\tau_z}
\end{array} \right),
\label{U}
\end{eqnarray}
transfers the phase $\phi_{0j}$ dependence from the diagonal term $\mathcal{H}_{jj}$ to the off-diagonal 
term $\mathcal{H}_{j,j+1}$ and to the wave function 
$|\Psi_j\rangle = \mathbb{U}_j |\psi_{0j}\rangle$. The tunneling amplitude between
$j$th and $(j+1)$th wires, which is proportional to $t_{\perp}$ in the off-diagonal Hamiltonian, acquires an
oscillating multiplier after transformation 
$\mathbb{H}_{j,j+1} = e^{i \frac{\phi_{0j}}{2}\tau_z} \mathcal{H}_{j,j+1} e^{-i \frac{\phi_{0j+1}}{2}\tau_z} 
\propto \frac{t_{\perp}}{2}\cos \frac{\phi_{j+1}-\phi_j}{2}~\tau_z$. Then, the  coupling energy 
$\Delta E_{j,j+1}$ due to the tunneling between two nearest-neighboring wires is 
$\Delta E_{j,j+1} \propto t_{\perp} \cos \frac{\phi_{j+1}-\phi_j}{2}$, yielding   
\begin{equation}
j_{j,j+1}=j_0^S(t_{\perp}) \sin \frac{\phi_{j+1}-\phi_j}{2}
\label{current}
\end{equation}
for the Josephson current. Note that each wire is chosen rather long, which ensures a negligible
overlapping of Majorana wave functions at two ends of a single wire. Nevertheless, Majoranas with
significant spatially overlapping wave functions in the nearest-neighboring wires survive \cite{dst15} due to 
a spatial reflection symmetry.
Tunneling of Majorana quasi-particles between the end of $j$th and $(j+1)$th wires 
results in Josephson current, flowing along the wires' end points. 
 
\emph{Conclusions}~~
In this paper we argue that a TRI topological superconducting phase may realizes in
the novel class of materials, consisting of regular, weakly-coupled superconducting wires in dielectric matrices 
\cite{bogom71, bk75, bogom78}.
The structures are fabricated under high pressures, which guarantee higher value of spin-orbit interactions.
Experimentally observed enhancement of the critical temperature of these structures allows us to suggest that apart 
from the intra-wire s-wave pairing, the inter-wire d-wave pairing sets up too, yielding an effective nodal order parameter.
The order parameter changes sign by crossing the nodal point between two Fermi surfaces, splitted due to
spin-orbit interactions.
Note that the only requirement for realization of a non-trivial topological superconductor is that
the superconducting pair potential switches sign between the two Fermi surfaces.  
Time-reversal symmetric topological superconductor belongs to a $DIII$ symmetry-class and is classified
by the $\mathbb{Z}_2$ topological invariant. 

The distance between the superconducting wires in the cavities or channels of the dielectric matrices is enough
large which allows us to detect the position of each wire under scanning tunneling microscope (STM). Topological phase
in a single wire can be manipulated by supplying an electrical potential by means of the cantilever of the STM,
which derives the wire from the superconducting phase to a normal metallic phase destroying Majorana quasi-particle
in a particularly chosen wire.  

\emph{Acknowledgments}--~~
S. M. kindly acknowledges support from  Research Grant EIF-2012-2(6)-39/08/1 of the Science Development 
Foundation under the President of the Republic of Azerbaijan.

\end{document}